\documentstyle[aps,prl,multicol,psfig]{revtex}

\begin{document}

\title{Micro-SQUID characteristics}
\author{K. Hasselbach}
\address{ CRTBT-CNRS, 25 avenue des Martyrs, BP 166 X, 38042 Grenoble, France}
\author{D. Mailly}
\address{LPN-CNRS, 196 H. Ravera, 92220 Bagneux, France} \author{J.R.
Kirtley} \address{ IBM T.J. Watson Research Center, Yorktown Heights,
NY 10598, USA} \date{\today} \maketitle
\begin{abstract}
We report on the dependence
on field and temperature
of the critical current of micro-SQUIDs: SQUIDs
with diameters as small as 1 micron, using
Dayem bridges as weak links. We model these SQUIDs by
solving the Ginzburg-Landau equations with appropriate boundary conditions
to obtain the supercurrent-phase relationships. These solutions show that
the phase drops and depression of the order parameter produced by
supercurrent flow are often distributed
throughout the micro-SQUID structure, rather than being localized in the
bridge area, for typical micro-SQUID geometries and coherence lengths.
The resultant highly
non-sinusoidal current-phase relationships $I_c(\varphi)$ lead to
reduced modulation depths and triangular
dependences of the micro-SQUID critical
currents on applied magnetic flux $I_c(\Phi)$.
Our modelling
agrees well with our measurements on both Al and Nb micro-SQUIDs.
\end{abstract}
\pacs{73.40.Gk,73.40.Rw,74.50.+r,74.72.-h}

\begin{multicols}{2}
\narrowtext

\section*{ INTRODUCTION}
Several different types of specialized Superconducting Quantum Interference
Devices (SQUIDs) have been developed for measuring the magnetic response
of small samples. One such device is the ``micro-SQUID":
a thin film DC-SQUID with Dayem-Bridges
as Josephson Junctions \cite{Chapelier_91}. In the micro-SQUID,
the entire device is fabricated by electron beam lithography, and the SQUID
loop itself serves as the flux-input coil. Micro-SQUIDs have the advantages
of very small pickup areas (about 1$\mu$m$^2$), and relatively small
sensitivity to in-plane applied magnetic fields (since they can be made
very thin, less than 20 nm). Micro-SQUIDs have been used for the observation
of persistent currents in 2 dimensional electron gas rings \cite{Chapelier_cp},
for the study of
the mechanisms of magnetization reversal in ferromagnetic particles as small as
3 nm in diameter \cite{Wernsdorfer_97,Wernsdorfer_01}, and have been
integrated into a scanning SQUID-AFM  with
high magnetic field spatial resolution \cite{Hasselbach_2000,Hasselbach_2001}. However, these
SQUIDs have two disadvantages: they have hysteretic current-voltage
characteristics, and relatively low modulation depths in their critical current -
flux characteristics. In this paper we report detailed measurements of
micro-SQUID characteristics, and compare them with model calculations.
These calculations model the micro-SQUID characteristics well, and could
represent a valuable tool for optimizing the properties of this class of
SQUIDs.

\section*{ Measurement}
The basic operating properties of micro-SQUIDs are understood:
They have a hysteretic V(I) characteristic, induced by the
propagation of a hot spot. As the current is ramped up from zero,
the micro-SQUID transits from the superconducting to the normal state at
a critical current I$_{c}$. A voltage step is generated as the
normal state resistance of
the junction appears and the dissipated energy heats the entire
micro-SQUID loop.
When the current is lowered the micro-SQUID stays in the resistive state until
the current is much smaller than I$_{c}$.
This thermal hysteresis excludes the usual current biasing schemes used
for DC-SQUID readout. Therefore a detection technique suitable for
hysteretic devices \cite{Fulton-72}
is implemented:
A computer controlled circuit simultaneously triggers a current ramp
and a 40MHz quartz clock. As soon as a $\partial V / \partial t$  pulse
of a preset height is detected at the micro-SQUID, the clock stops and the
current is set to zero.
The clock reading is transferred to the computer, and the cycle begins again.
The critical current is proportional to the duration of the current ramp.
The fastest repetition rate is 10kHz, limited by the time needed to
settle the current. A single wire is sufficient to connect the
micro-SQUID, since
the  $\partial V / \partial t$  pulse is detected on the current
biasing lead.
Every time the critical current is measured, the flux state in the
micro-SQUID is sampled, and every time the micro-SQUID
becomes normally conducting
the external field penetrates. As the current is reduced (in 40ns) to zero
the micro-SQUID structure becomes progressively superconducting again and
screening currents are set up to quantize the total flux
through the micro-SQUID.
In the limit of high critical currents different flux configurations
can be
stabilized in the micro-SQUID ring during the backswitching, leading to
multivalued I$_c$ vs B characteristics.

However, the details of the dependence of the critical current of the
micro-SQUID on magnetic field and temperature have not been well characterized.
We present in this manuscript a detailed description of the
underlying physics necessary for the understanding and improvement of this type
of SQUID. In this study
we made I$_c$(T) and I$_c$ vs B measurements on micro-SQUID devices made
of aluminum or niobium. Each SQUID consists of a 1$\mu$m  square loop.
Bulk aluminum and niobium  have very different superconducting
properties, such as the coherence length $\xi$,
and the normal to superconducting transition temperature T$_c$.
The characteristics of micro-SQUIDs (see e.g. Fig.'s 2 and 4)
do not resemble those of ideal
Josephson SQUIDs, in that they have relatively shallow modulation depths and
triangular I$_c$ vs B interference patterns at low temperatures.

We believe that these non-ideal characteristics occur because
the weak links in these SQUIDs are Dayem-Bridges, with dimensions
comparable to the coherence length.
It is well known that ``long" Josephson
weak links can have non-sinusoidal Josephson
current-phase relationships \cite{likharev76a,likharev76b,harris},
and that non-sinusoidal Josephson current-phase relationships can in
turn lead to non-ideal SQUID I$_c$ vs B interference patterns\cite{vanduzer}.
Faucher et al.\cite{faucher} have attributed the non-ideal critical
current vs. flux characteristics of their micro-SQUIDs to the large
kinetic inductance of the micro-bridges. We believe the treatment given
here is equivalent to that of Faucher et al. in terms of the properties
of the micro-bridges, but it goes further, in that it includes the properties
of the entire SQUID loop itself.
We model the SQUID current-phase relationships by solving the 2-d
Ginzburg-Landau equations. It is important to solve these equations
in the full geometry of the micro-SQUID, since as
previously pointed out in general for Dayem-Bridges\cite{likharev76b},
and as is born out by our
modelling in this particular case, the
phase drops and order-parameter depression associated with supercurrent
flow often extend far from the micro-bridge.

\section*{ Model}

The first Ginzburg-Landau (GL) differential equation is\cite{tinkham}
\begin{equation}
\alpha \psi + \beta \mid \psi \mid ^2 \psi + \frac{1}{2m^*} \left (
\frac{\hbar}{i} \vec{\nabla} - \frac{e^* \vec{A}}{c} \right ) ^2 \psi = 0
\label{eq:gldiff}
\end{equation}
where $m^*=2m$ and $e^*=2e$ are the mass and charge of the Cooper pair,
$\psi$ is the complex order parameter describing the superconducting
state, and $\vec{A}$ is the vector potential. The supercurrent $\vec{J}$
is given by:
\begin{equation}
\vec{J} = \frac{e^* \hbar}{2m^* i} (\psi^* \vec{\nabla}\psi - \psi
\vec{\nabla} \psi ^* ) - \frac{e^{*2}}{m^*c} \psi^*\psi\vec{A}.
\label{eq:glsup}
\end{equation}
For the purposes of the paper, we include the effects of the vector
potential $\vec{A}$ in a lumped circuit element model (see Eq. \ref{eq:phisum}).
Therefore in our
solution of the GL equations we set $\vec{A}=0$. Writing the
complex order parameter as $\psi = \mid \psi \mid e ^{i \varphi}$, using
a coordinate system in which the SQUID is in the $xy$ plane,
and neglecting the $z$ dependence of the gradients of $\psi$, the
real (Eq. \ref{eq:glreal})
and imaginary (Eq. \ref{eq:glimag}) parts of Eq. \ref{eq:gldiff} become:
\begin{eqnarray}
\alpha \mid \psi \mid + \beta \mid \psi \mid ^3 - \frac{\hbar ^2}{2 m*}
\left [ \frac{d^2 \mid \psi \mid}{dx^2} - \mid \psi \mid \left (\frac{d\varphi}
{dx} \right )^2 \right ] \nonumber \\
- \frac{\hbar ^2}{2m*}
\left [ \frac{d^2 \mid \psi \mid}
{dy^2} - \mid \psi \mid \left ( \frac{d \varphi}{dy} \right )^2 \right ] = 0
\label{eq:glreal}
\end{eqnarray}
and
\begin{equation}
2 \frac{d\varphi}{dx}\frac{d \mid \psi \mid}{dx}+\mid \psi \mid \frac{d ^2 \varphi}{dx^2}
+ 2 \frac{d\varphi}{dy}\frac{d \mid \psi \mid}{dy}+\mid \psi \mid \frac{d ^2 \varphi}{dy^2}=0.
\label{eq:glimag}
\end{equation}
Setting $\mid \psi \mid = f \mid \psi \mid _{\infty}$, where $f$ is a real
function of $x$ and $y$, and $\mid \psi \mid _{\infty}$ is the unperturbed
value of the order parameter, $\beta \mid \psi \mid ^2 _{\infty} =
-\alpha = \hbar ^2 /2 m ^* \xi ^2 (T)$ \cite{tinkham}, and using
reduced units $x'=x/\xi (T), y'=y/\xi (T)$,
Eq. \ref{eq:glreal}
becomes
\begin{equation}
\frac{d^2f}{dx'^2}-f\frac{d^2\varphi}{dx'^2}
+\frac{d^2f}{dy'^2}-f\frac{d^2\varphi}{dy'^2}
+f-f^3=0
\label{eq:glreal2}
\end{equation}
Eq. \ref{eq:glimag} becomes
\begin{equation}
2 \frac{df}{dx'}\frac{d\varphi}{dx'}+f \frac{d^2\varphi}{dx'^2}
+2 \frac{df}{dy'}\frac{d\varphi}{dy'}+f \frac{d^2\varphi}{dy'^2}=0.
\label{eq:glimag2}
\end{equation}
The supercurrent density $\vec{J}$ becomes
\begin{equation}
\vec{J} = \frac{e^* \hbar}{m^* \xi(T)} \mid \psi \mid _{\infty} ^2 f ^2
\left [ \hat{x} \frac{d\varphi}{dx'}+ \hat{y} \frac{d\varphi}{dy'} \right ].
\label{eq:glsup2}
\end{equation}
It can be shown that Eq. \ref{eq:glimag2} is equivalent to setting the divergence of the
supercurrent (Eq. \ref{eq:glsup2}) equal to zero.

\begin{figure}
\centerline{\psfig{figure=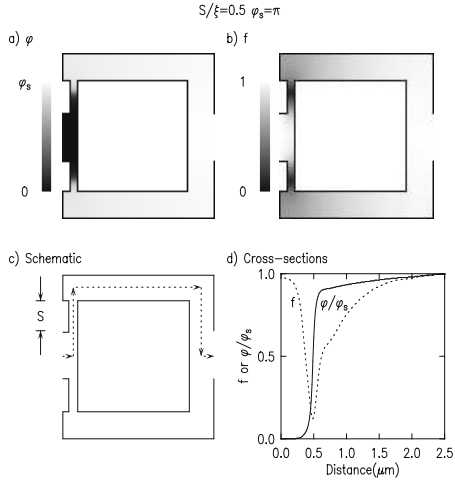,height=2.5in}}
\vspace{0.1in}
\caption{Solution to the 2-dimensional Ginzburg-Landau differential equations
for the quantum mechanical phase $\varphi$ and normalized superfluid
density $f$ for a micro-SQUID. The geometry of the SQUID is that of the Al
micro-SQUID for which measurements are presented in Fig. \ref{fig:musquial}.
The length of the bridge $S=290$nm,
bridge width=66 nm, the SQUID length (horizontal,$x$) is 1.3$\mu$m  and
width (vertical,$y$) is 1.45$\mu$m. The calculations for this figure were done
for $\xi$=580nm, with a total phase drop across the SQUID of $\varphi_s=\pi$.
Greyscale images of the superconducting phase $\varphi$ (a) and
superfluid density $f$ (b) in the SQUID are shown. Shown in (d) are cross-sections
through these images, along the path shown in (c). For this value of
$S/\xi(T)$=0.5, the phase drop and superfluid depression are
mostly localized in the micro-bridge region.}
\label{fig:glsquid3}
\end{figure}

By taking discrete steps $\delta$ in both the $x$ and $y$ directions,
the differential equation Eq. \ref{eq:glreal2} can be cast as a difference
equation
\begin{eqnarray}
f_{i+1,j}+f_{i-1,j}+f_{i,j+1}+f_{i,j-1}
+(-4-(\varphi_{i+1,j}-\varphi_{i,j})^2 \nonumber \\
-(\varphi_{i,j+1}-\varphi_{i,j})^2 + \delta ^2)f_{i,j}=\delta ^2 f_{i,j}^3,
\label{eq:gldif1}
\end{eqnarray}
where $i$ and $j$ are indices labelling the 2-d matrices in the $x$ and $y$
directions respectively.
Similarly, Eq. \ref{eq:glimag2} becomes
\begin{eqnarray}
\varphi_{i+1,j} \left [ f_{ij} + 2 ( f_{i+1,j}-f_{i,j} ) \right ]
+\varphi_{i-1,j}f_{i,j}  \nonumber \\
+ \varphi_{i,j+1} \left [ f_{i,j} + 2 ( f_{i,j+1} - f_{i,j} )
\right ] + \varphi_{i,j-1} f_{i,j} \nonumber \\
+ \varphi_{i,j} \left [ -4 f_{i,j} -2 ( f_{i+1,j}
-f_{i,j}) - 2 ( f_{i,j+1} - f_{i,j} ) \right ] = 0.
\label{eq:gldif2}
\end{eqnarray}
Using (in MKS units)
$\mid \psi \mid ^2 _{\infty} = m/2 \mu_0 e^2 \lambda_{eff}^2$, where
$\lambda_{eff}$ is the
effective thin film penetration depth\cite{tinkham},
the total current through the SQUID can be written  as:
\begin{equation}
I = I_0 \sum _j f_{i,j}^2(\varphi_{i+1,j} -\varphi_{i,j}),
\label{eq:cursum}
\end{equation}
where $I_0=d \hbar /2 \mu_0 e \lambda_{eff}^2$,
and $d$ is the thickness
of the film from which the
micro-SQUID is patterned. Supercurrent conservation requires that the
sum over $j$ in Eq. \ref{eq:cursum} be independent of the value of $i$
chosen. This was used as a self-consistency check of the solutions presented
below. The difference equations Eq. \ref{eq:gldif1},\ref{eq:gldif2} are in
the standard form
\begin{equation}
a_{j,l}u_{j+1,l}+b_{j,l}u_{j-1,l}+c_{j,l}u_{j,l+1}+d_{j,l}u_{j,l-1}+
e_{j,l}u_{j,l} = g_{j,l}.
\label{eq:standdif}
\end{equation}
We solved Eq.'s \ref{eq:gldif1} and \ref{eq:gldif2}
using a general non-linear differential equation
solver subroutine (SOR) using the
successive over-relaxation method with Chebyshev acceleration\cite{press}.
An initial guess was made for the matrices $f_{i,j}$ and $\varphi_{i,j}$,
SOR was used to search for a solution for $f_{i,j}$ of
Eq. \ref{eq:gldif1} for a pre-determined number of iterations,
with $\varphi_{i,j}$ fixed. The result
for $f_{i,j}$ was then fixed, and a solution for $\varphi_{i,j}$ in
Eq. \ref{eq:gldif2} was
sought for the same number of iterations. This
procedure was iterated until
the sum of the absolute values of the deviations from equality in
Eq.'s \ref{eq:gldif1} and \ref{eq:gldif2} were both less
than a fixed error sum value. For the results reported here, the error sum value
was chosen to be 10$^{-3}$. This procedure gave results in agreement with
those reported by Likharev and Yakobson for long 1-D microbridges\cite{likharev76a}.
For modelling of our micro-SQUIDs we chose
the boundary conditions:
1) $\varphi$ was fixed at $0$ along the entrance to the micro-SQUID
structure (at the left of Fig. \ref{fig:glsquid3}a)), and fixed at
$\varphi_s$ at the
exit of the micro-SQUID structure (to the right);
2) $f$ was chosen to be 1 at both the
entrance and exits to the micro-SQUID; and 3) the components of the
gradients of $f$ and $\varphi$ were taken to be zero normal to the other
boundaries (solid lines in Fig. \ref{fig:glsquid3}a,b). The boundaries of
the model calculations were chosen to match those of electron micrographs
of the actual SQUIDs measured. In choosing these boundary conditions
we neglect the effects
of phase drops and supercurrent depression in the leads to and from
the micro-SQUID.
We argue that phase drops outside of the SQUID loop should have little
effect on the critical current-flux characteristics which we are modelling,
and that neglect of lead effects is therefore a good starting approximation.

\begin{figure}
\centerline{\psfig{figure=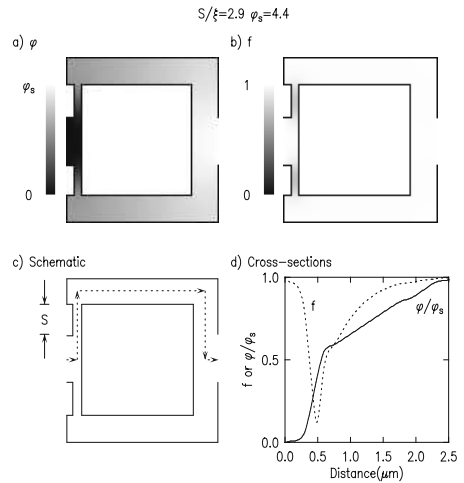,height=2.5in}}
\vspace{0.1in}
\caption{Calculations for the same geometry as Fig. \ref{fig:glsquid3}, but
in this case for S/$\xi$=2.9, with a total phase drop $\varphi_s=$4.4.
In this case both the phase drop and superfluid density depression extend
throughout the SQUID structure.}
\vspace{0.1in}
\label{fig:glsquid4}
\end{figure}

Example solutions for $\varphi$ and $f$ using this model are shown in
Fig.'s \ref{fig:glsquid3} and \ref{fig:glsquid4}. Here the geometry is that
of the Al micro-SQUID. The $\varphi_{i,j}$ and $f_{i,j}$ matrices both had
102$\times$110 elements, with 76 elements/$\mu$m. The bridges were $S$=289 nm
long and 66 nm wide. The arms of the SQUID were 237 nm wide.
Figure \ref{fig:glsquid3} shows the solution to Eq.'s \ref{eq:gldif1} and
\ref{eq:gldif2} for $S/\xi$=0.5, $\varphi_s=\pi$. For a ``short" 1-d micro-bridge
($S/\xi << $ 1) with $\varphi_s=\pi$,
the solution for $\varphi$ has a step from $0$ to $\pi$, and a
depression in $f$ to 0, with a characteristic width of $\xi$, centered at the
center of the bridge\cite{baratoff}. The numerical solution for the full SQUID
structure has a step in $\varphi$ and a depression in $f$ near the center of
the bridges, but there are significant changes well into the body of the SQUID,
and $f$ does not go completely to $0$, meaning that $I_s$ does not go to zero,
until $\varphi_s$ is slightly above $\pi$. These effects become more pronounced
as $S/\xi$ increases. Figure \ref{fig:glsquid4} shows the results for $S/\xi$=
2.9, $\varphi_s$=4.4.
This is the maximum value of $\varphi_s$ for which a
numerical solution to the Ginzburg-Landau equations
could be found for $S/\xi=$2.9.
In this case almost half of the total phase drop occurs outside of
the micro-bridge region.

\begin{figure}
\centerline{\psfig{figure=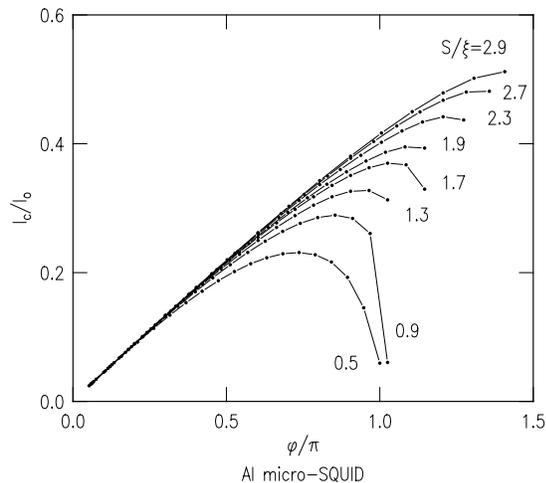,height=2.5in}}
\vspace{0.1in}
\caption{Calculated supercurrent-phase relationship for the Al SQUID, for
values of $S/\xi$ appropriate for the data
shown in Fig. \ref{fig:musquial}}.
\vspace{0.1in}
\label{fig:glalivp}
\end{figure}

\begin{figure}
\centerline{\psfig{figure=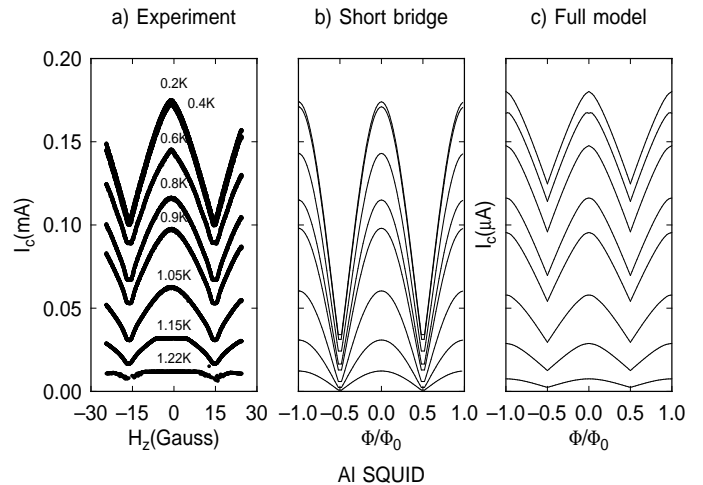,height=2.5 in}}
\vspace{0.1in}
\caption{(a) Experimental measurements of the critical current of an Al
micro-SQUID as a function of applied field, for selected temperatures.
(b) Modelling in the short-bridge limit I=I$_1$sin$\varphi$,
and (c) Modelling using the full Ginzburg-Landau calculations described
in the text.}
\vspace{0.1in}
\label{fig:musquial}
\end{figure}

Supercurrent-phase relationships $I_s(\varphi)$ for this geometry are shown in
Figure \ref{fig:glalivp}. This procedure is unable to find solutions numerically
for the lower branch of the $I_s(\varphi)$ characteristic when this characteristic
is doubly valued \cite{likharev76a,likharev76b}. This will not affect our
results, since we are comparing our modelling with the
maximum supercurrent at each value of applied field. This conclusion is
supported by our calculations of $I_c(\Phi)$ characteristics using the
Likharev-Yakobson \cite{likharev76a} $I_c(\varphi)$ characteristics: only
the upper branch is important for calculating the $I_c(\Phi)$ characteristics.
The coherence lengths $\xi$ in Fig. \ref{fig:glalivp} were chosen to be
appropriate for an Al micro-SQUID at the temperatures for which the measurements
of Fig. \ref{fig:musquial} were made: We use
the dirty limit
expression for the temperature dependent coherence length
$\xi(T) = 0.855 \sqrt{\xi_0 l/(1-T/T_c)}$ \cite{tinkham},
estimate the low temperature mean free
path $l$ in our films to be about 10nm from transport measurements, and
take a value for $\xi_0$ of 100nm \cite{maloney}, and $T_c$=1.25K.

A SQUID with non-sinusoidal current-phase ($I_c(\varphi)$)
relationships will have critical current vs. flux interference
patterns ($I_c(\Phi)$) that
are also non-standard. These can be modelled as follows. Assume that the
SQUID has two arms labelled $a$ and $b$, with total phase drops across the
two arms $\varphi_a$ and $\varphi_b$,
inductances $L_a$ and $L_b$,
and micro-bridge supercurrents $I_{sa}=I_a g(\varphi_a)$ and
$I_{sb}=I_b g(\varphi_b)$.

The requirement of a single valued superconducting order parameter
leads to
\begin{equation}
2 \pi n = \varphi_a - \varphi_b +2 \pi \phi_e + \beta_a g(\varphi_a) - \beta_b
g(\varphi_b),
\label{eq:phisum}
\end{equation}
where $\phi_e$ is the externally applied flux $\Phi_e$ divided by the
superconducting
flux quantum $\Phi_0=h/2e$, $\beta_{a} = 2\pi L_{a} I_{a}/\Phi_0$ and
$\beta_{b} = 2\pi L_{b} I_{b}/\Phi_0$.
The total supercurrent through the SQUID is
\begin{equation}
I = I_ag(\varphi_a)+I_bg(\varphi_b).
\label{eq:currsum}
\end{equation}

The dependence of the critical current on
applied field is determined most easily
by assuming values for $\phi_e$ and one of the micro-bridge phase differences
$\varphi_a$ or $\varphi_{b}$, then varying the phase of the second micro-bridge
until Eq. \ref{eq:phisum} is satisfied. The values for $\varphi_a$
and $\varphi_b$ are then substituted into Eq. \ref{eq:currsum} to find the
total current. The maximum value for $I$ after repeating this procedure
for all initial values of $\varphi_a$ and $\varphi_b$
is the critical current for that value of $\phi_e$.
The effect of the self-induced field \cite{Fulton-72} is expressed in
Eq. \ref{eq:phisum} via $\beta(T) = 2\pi L I_{c}(T)/\Phi_0$.
$\beta$(T) increases from $T_c$ in proportion to the critical
current, since the inductance is purely geometric.
In the case of the Al micro-SQUIDs, with a critical current of 170 $\mu$A
$\beta$(0.2 K) is 0.78; in the case of Nb, which have critical currents
of 2000 $\mu$A, $\beta$(0.2 K) can be as high as 8.5. However, the differences
in $I_c$ are not enough to explain the differences in the $I_c(\Phi)$
characteristics between Al and Nb: they also arise from the shorter
coherence length in Nb, which leads to more highly non-sinusoidal current-phase
relationships in micro-SQUIDs made from Nb.

Figure \ref{fig:musquial}a shows critical current vs. applied magnetic
field characteristics for our Al micro-SQUID.
Figure \ref{fig:musquial}b shows the prediction
of Eq. \ref{eq:phisum} and Eq. \ref{eq:currsum} assuming the standard
Josephson current-phase relationship I$_c$=I$_1$ sin($\varphi$) for the
micro-bridges. In this modelling the critical currents of the bridges and
the inductances of the arms of the SQUIDs were assumed to be symmetric and
the critical currents were chosen to match the zero field critical current
of the SQUID at each temperature. The inductances L$_a$ and L$_b$ were chosen
to be 0.76pH each, half of the calculated total inductance
$L=5\mu_0 C/16$, where $C$ is the inner circumference
of the SQUID \cite{jaycox}.

\begin{figure}
\centerline{\psfig{figure=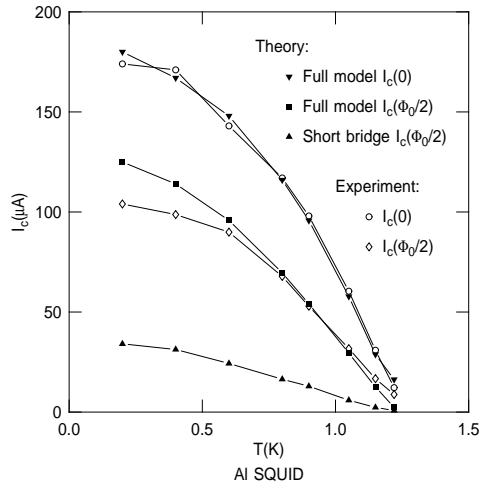,height=2.5 in}}
\vspace{0.1in}
\caption{Plot of the experimentally measured critical currents at zero
field (open circles) and at an applied flux of half the full modulation
period (open diamonds), for an Al micro-SQUID, as a function of temperature.
The solid triangles are modelling as described in the text for the
conventional sinusoidal current-phase relationship. The solid inverted triangles
and squares are
modelling using the Ginzburg-Landau calculations described in the text.}
\label{fig:glalmod}
\end{figure}

Figure \ref{fig:musquial}c shows the predictions
using the full GL calculations for the current-phase relationship,
with the same values for the inductances.
For this modelling we take
$I_{ca}(\varphi_a)=I_{ca}(\varphi_b)=I_c(\varphi)/2$, where $I_c(\varphi)/I_0$
is shown for the Al micro-SQUID
in Fig. \ref{fig:glalivp}. In the determination of $I_0$ we assume
$\lambda^2_{eff}(T)=\lambda^2_{eff}(0)/1-t^4$, $t=T/T_c$, $d$=38nm
and use $\lambda_{eff}(0)$ as the sole fitting parameter. The best fit between
experiment and modelling for $I_c(T,\Phi = 0)$
is for $\lambda_{eff}(0)$=172 nm.
This is to be
compared with $\lambda_L$ = 44 nm for bulk aluminum \cite{maloney}.
The effective penetration depth is often longer in thin films than in bulk,
and can also be increased by impurity scattering \cite{tinkham,maloney}.
Further, the effective thickness of the films will be reduced by oxidation.
Once this scaling is done, the agreement between theory and experiment
for $I_c(T,\Phi=0)$ is good.
The full model
(Figure \ref{fig:musquial}c) fits the $I_c(T,\Phi)$ experimental data well.
This is made clear in
the plot of $I_c(0)$, the critical current at zero applied field, and
$I_c(\Phi_0/2)$, the critical current at the first minimum of the interference
pattern, as a function of temperature, in Figure \ref{fig:glalmod}.
In this figure, the open circles are $I_c(0)$, and the open diamonds
are $I_c(\Phi_0/2)$. The solid triangles are the predictions for a
sinusoidal current-phase relationship model as outlined above. The
solid squares and inverted triangles are the predictions using the
GL current-phase relationship described above.

\begin{figure}
\centerline{\psfig{figure=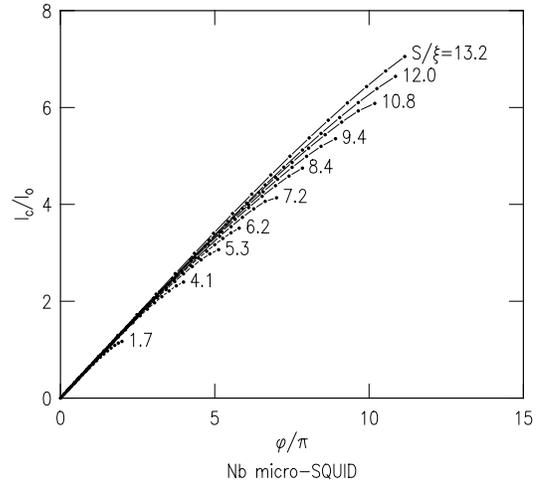,height=2.5in}}
\vspace{0.1in}
\caption{Calculated supercurrent-phase relationship for the Nb SQUID, for
various values of $S/\xi$ appropriate for the data
at selected temperatures
shown in Fig. \ref{fig:musquinb}}.
\label{fig:glnbivp}
\end{figure}

\begin{figure}
\centerline{\psfig{figure=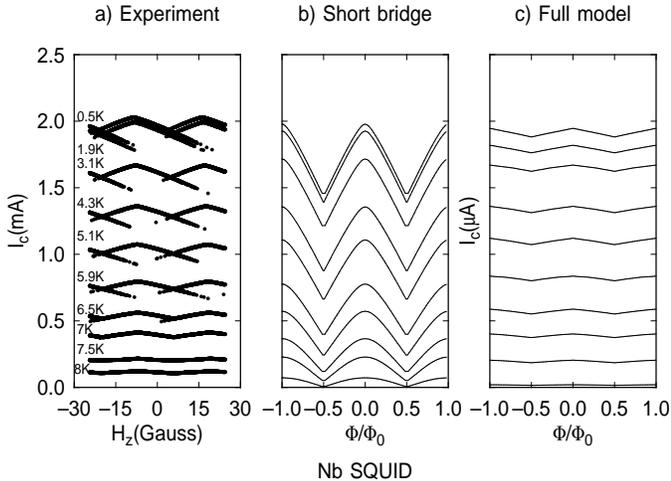,height=2.5in}}
\vspace{0.1in}
\caption{(a) Experimental measurements of the critical current of an Nb
micro-SQUID as a function of applied field, for selected temperatures.
(b) Modelling in the short-bridge limit I=I$_1$sin$\varphi$,
and (c) Modelling using the full Ginzburg-Landau calculations described
in the text.}
\label{fig:musquinb}
\end{figure}

Similar conclusions can be drawn from the data and modelling for a Nb
micro-SQUID (Fig. \ref{fig:musquinb}). In this case the $f_{i,j}$ and
$\varphi_{i,j}$ matrices were 100$\times$100, with 67 elements/micron.
The micro-bridges were 184 nm long and 100 nm wide, with the arms of the
SQUID 285 nm wide. $I_c(\varphi)$ characteristics for the values of
$S/\xi(T)$ appropriate for the data of Fig. \ref{fig:musquinb} are shown
in Fig. \ref{fig:glnbivp}. In this case $S/\xi(T)$ is large, the variations
of $\varphi$ and $f$ are spread throughout the SQUID structure,
and $I_c(\varphi)$
is nearly linear, at all temperatures. For the calculations of the $I_c(\Phi)$
characteristics in Fig. \ref{fig:musquinb}c,
the total inductance of the SQUID was taken to be
1.4 pH, the dirty limit expression
$\xi(T)=0.855(\xi_0 l)^{1/2}/(1-t)^{1/2}$
was again used for the coherence length, with
$l$=6.5nm \cite{harris},
$\xi_0=39nm$ \cite{harris},
$\lambda_{eff}^2(T)=\lambda_{eff}^2(0)/1-t^4$, $d$=30nm,
and $T_c$ = 8.23K.
Figure \ref{fig:musquinb}a shows experimental measurements
of the dependence of the SQUID critical current on applied magnetic field
at selected temperatures.
A multivalued critical current appears for currents higher than
0.4 and 0.5 mA for our given inductance.  The sections of
of the $I_c(\Phi)$ characteristics are nearly linear, with sharp
discontinuities in the slopes at $\Phi=(n+1/2)\Phi_0$, n an integer.

\begin{figure}
\centerline{\psfig{figure=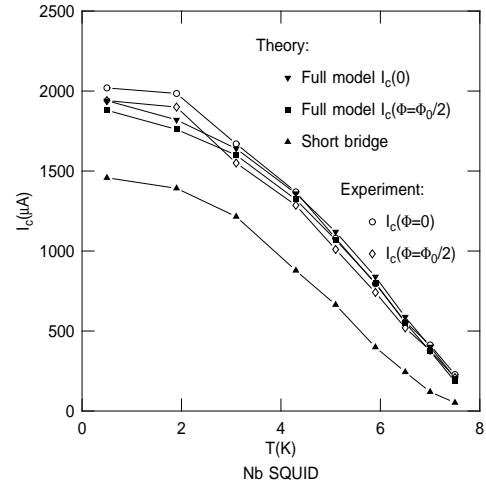,height=2.5in}}
\vspace{0.1in}
\caption{Plot of the experimentally measured critical currents at zero
field (open circles) and at an applied flux of half the full modulation
period (open diamonds), for a Nb micro-SQUID, as a function of temperature.
The solid triangles are modelling as described in the text for the
conventional sinusoidal current-phase relationship. The solid inverted triangles
and squares are
modelling using the full Ginzburg-Landau calculations for the current-phase
relationship, as described in the text.}
\label{fig:glnbmod}
\end{figure}

Figure \ref{fig:musquinb}b shows modelling for
an assumed sinusoidal current-phase relationship, using symmetric
experimentally determined values for the micro-bridge critical currents
and symmetric inductances of 0.7pH for each arm of the SQUID. As for the
Al micro-SQUID case, this modelling does poorly in describing the
$I_c(\Phi)$ characteristics.
Figure \ref{fig:musquinb}c shows modelling using the GL calculations
expressions for the micro-bridge current-phase relationship, with
symmetric inductances of 0.7pH
in each arm, and a best fit value of $\lambda_{eff}(0)$= 173 nm. This is
to be compared with $\lambda_L(0)$= 44 nm for bulk Nb with a T$_c$ of 9.26K
\cite{barone}. As can be seen from Figures \ref{fig:musquinb}c and
\ref{fig:glnbmod}, the full modelling predicts the temperature dependence
of the SQUID critical current well, shows a triangular
$I_c(\Phi)$ dependence, and
does significantly better than that using a sinusoidal
current-phase relationship for the modulation depth.

\section*{ Discussion}

Our data shows that the
modulation depth is reduced compared to the short bridge model
as soon as the coherence length becomes shorter than the bridge length.
This effect is much more pronounced in the case of Nb as the
intrinsic short coherence length of Nb leads rapidly to values of
S/$\xi$ larger than 1.
The model describes very well the overall lineshape in the case of both the
Al micro-SQUIDs, as well as the Nb micro-SQUIDs, which show pronounced
triangular $I_c(\Phi)$ characteristics. It is remarkable that the
Nb micro-SQUIDs show quantum interference at all temperatures, even when the
micro-bridges are  much longer than the coherence length, and the phase
gradients and supercurrent density depressions extend throughout the
body of the SQUID. This implies that the heat dissipation that takes
place when the SQUID enters the voltage state occurs well outside of
the micro-bridge regions.

How can the sensitivity of these micro-SQUIDs be improved?
The Nb SQUIDs are limited in their sensitivity by the short coherence
length of Nb, as the modulation depth is maximal if the coherence length
is equal to or
longer than the bridge length. The coherence length could be
increased by increasing the
mean free path (epitaxial films or thicker layers), but an increase in
critical current must be avoided. The performance of these micro-SQUIDs
could be improved by significantly reducing the length of the
Dayem bridges, perhaps using novel
scanning probe techniques \cite{bouchiat}.
Decreasing thickness and width of
the micro-bridge may lead to a diminution of the mean free path and
thus to a diminution of the coherence length.
A further way to increase sensitivity of Al as well as Nb micro-SQUIDs may be
the suppression of thermal hysteresis.
A non-hysteretic behavior may be attained by fabricating the entire SQUID
on a normal metal plane in order to remove the heat produced by the
non local phase relaxation in the microSQUID. The non-hysteretic
behavior will enhance the intrinsic bandwidth and allow for a standard
DC-SQUID detection scheme.

\section*{ Conclusion}
In summary, the temperature and field dependence of the critical currents
in micro-SQUIDs can be
understood by means of numerical calculations based on the phase
dependence of the critical current predicted by a 2-d Ginzburg-Landau
numerical calculation. It is important to do this calculation for the
full micro-SQUID structure in many cases, because of the spreading
of the variation in the superconducting phase and depression in the
supercurrent density beyond the micro-bridge region in this type of
device.

\section*{ Acknowledgments:}
 J. K. would like to thank F. Tafuri for useful conversations, and
the Universit$\acute{e}$
Joseph Fourier of Grenoble for their
support during his visit in 1998,
and A. Benoit is thanked for his encouragement of this work.

\end{multicols}
\end{document}